\documentclass[manuscript,screen]{acmart}

\AtBeginDocument{%
  }

\copyrightyear{2024} 
\acmYear{2024} 
\setcopyright{acmlicensed}\acmConference[LAK '24]{The 14th Learning Analytics and Knowledge Conference}{March 18--22, 2024}{Kyoto, Japan}
\acmBooktitle{The 14th Learning Analytics and Knowledge Conference (LAK '24), March 18--22, 2024, Kyoto, Japan}
\acmPrice{15.00}
\acmDOI{10.1145/3636555.3636898}
\acmISBN{979-8-4007-1618-8/24/03}

\usepackage{subcaption}
\usepackage{enumitem}
\usepackage{pifont}
\usepackage{pgf-pie}
\usepackage{pgfplots}
\usepackage{booktabs} 
\usepackage{multirow}

\begin{document}

\title{Finding Paths for Explainable MOOC Recommendation: A Learner Perspective}


\author{Jibril Frej}
\affiliation{%
  \institution{EPFL}
  \city{Lausanne}
  \country{Switzerland}}
\email{jibril.frej@epfl.ch}

\author{Neel Shah}
\affiliation{%
  \institution{EPFL}
  \city{Lausanne}
  \country{Switzerland}}
\email{neelshah29042002@gmail.com}

\author{Marta Knežević}
\affiliation{%
  \institution{EPFL}
  \city{Lausanne}
  \country{Switzerland}}
\email{marta.knezevic@epfl.ch}

\author{Tanya Nazaretsky}
\affiliation{%
  \institution{EPFL}
  \city{Lausanne}
  \country{Switzerland}}
\email{tanya.nazaretsky@epfl.ch}

\author{Tanja Käser}
\affiliation{%
  \institution{EPFL}
  \city{Lausanne}
  \country{Switzerland}}
\email{tanja.kaeser@epfl.ch}







\renewcommand{\shortauthors}{Jibril Frej et al.}
\newcommand{\red}[1]{{\color{red}#1}}
\newcommand{\gray}[1]{{\color{gray}#1}}

\begin{abstract}
The increasing availability of Massive Open Online Courses (MOOCs) has created a necessity for personalized course recommendation systems.
These systems often combine neural networks with Knowledge Graphs (KGs) to achieve richer representations of learners and courses. 
While these enriched representations allow more accurate and personalized recommendations, explainability remains a significant challenge which is especially problematic for certain domains with significant impact such as education and online learning.
Recently, a novel class of recommender systems that uses reinforcement learning and graph reasoning over KGs has been proposed to generate explainable recommendations in the form of paths over a KG.
Despite their accuracy and interpretability on e-commerce datasets, these approaches have scarcely been applied to the educational domain and their use in practice has not been studied.
In this work, we propose an explainable recommendation system for MOOCs that uses graph reasoning.
To validate the practical implications of our approach, we conducted a user study examining user perceptions of our new explainable recommendations.
We demonstrate the generalizability of our approach by conducting experiments on two educational datasets: \texttt{COCO} and \texttt{Xuetang}.
\end{abstract}




\maketitle


\section{Introduction}

The proliferation of Massive Open Online Courses (MOOCs) has led to a democratization of educational resources, yet it has also introduced an information overload problem. To illustrate, Udemy provides over 213,000 courses, including 10,500 accredited ones, while Coursera hosts over 14,000 courses, with 7,000 being accredited. This overwhelming variety highlights the essential role of effective recommendation systems in assisting learners in selecting from the myriad of available courses. These systems are essential in helping learners find the most suitable courses based on their individual needs (e.g., goals, backgrounds, and motivations). They can facilitate optimal learning experiences and play a pivotal role in effectively steering learners' academic and professional paths. Indeed, for a recommendation to be truly impactful, it must be tailored to address the students' diverse learning objectives, skill levels, and aspirations. 

Recent advancements in neural network-based recommender systems have set new standards for generating precise and individualized course suggestions~\cite{DBLP:journals/jbd/RoyD22}. Nonetheless, the majority of these models serve as black boxes leaving the rationale behind their recommendations opaque. This lack of transparency can diminish learners' trust and their willingness to accept the suggested recommendations~\cite{DBLP:conf/chi/Kizilcec16}, highlighting the tradeoff between model accuracy and interpretability. Given the significant impact of educational choices, and considering the proven connection between clear, understandable recommendations and trust among learners~\cite{DBLP:conf/chi/Kizilcec16}, there exists a clear need for algorithms that are not only accurate but also explainable. Such algorithms should explain their suggestions, assisting learners in making well-informed decisions by balancing accuracy with clarity and transparency in the recommendation process.


In a variety of domains, explainable recommendation systems have garnered considerable attention as an active area of research~\cite{DBLP:journals/ftir/ZhangC20}. Different approaches have been explored, including factorization models which explain recommendation by selecting item features from user reviews~\cite{DBLP:conf/sigir/ZhangL0ZLM14} and topic modeling approaches that provides users with topic word clouds~\cite{7113324,10.1145/2684822.2685291}. Graph-based models and knowledge-graph-based explanations have been developed to generate sentences explaining the recommendation that uses relations in the knowledge graph (KG)~\cite{DBLP:journals/corr/abs-1710-07134}. Additional strategies range from leveraging the attention mechanism~\cite{DBLP:conf/nips/VaswaniSPUJGKP17} to select item reviews serving as explanations~\cite{DBLP:conf/www/ChenZLM18} and employing modern Large Language Models to generate explanations in natural language~\cite{DBLP:journals/corr/abs-2309-08817, DBLP:conf/recsys/ChangHT16}. Nonetheless, the majority of these explanations are generated post-hoc and may not accurately represent the model's underlying reasoning.


Recent advancements in Reinforcement Learning (RL) applied to KG reasoning for recommendation offer intrinsic interpretability without compromising predictive performance. In this paradigm, an RL agent navigates through the KG using relations between entities, starting from a learner and concluding at the course to be recommended, thereby inherently providing an interpretable line of reasoning. To our knowledge, Policy-Guided Path Reasoning (PGPR)~\cite{Xian_2019} was the first approach to use RL applied to KG reasoning for explainable recommendation. Since PGPR, several improvement have been proposed ~\cite{song2019ekar, multi-level-reasoning, dialogue_mining, DBLP:conf/cikm/XianFZGCHG0MMZ20}.


In the educational domain, course recommendation systems have been the subject of extensive study, covering a wide variety of aspects. Investigations have been conducted into serendipity-based diverse course recommendation~\cite{DBLP:conf/lak/PardosJ20} and explainable learning activities recommendation through open learner models (OLMs)~\cite{DBLP:conf/lak/AbdiKSG20}. Research has also delved into peer learner recommendation~\cite{10.1145/3170358.3170400}, target course-oriented recommendation~\cite{10.1145/3303772.3303814}, and the recommendation of short video clips to mitigate information overload~\cite{2022.EDM-posters.86}. In the specific domain of MOOCs recommender systems using neural networks (NN), multiple research directions have been pursued. These include optimizing the accuracy of recommendation~\cite{SanguinoPerez_Manrique_Mariño_LinaresVásquez_Cardozo_2022, sakboonyarat2019massive, khalid2021novel, 10.1145/3303772.3303814, 2022.EDM-posters.86}, ensuring fairness~\cite{10.1145/3404835.3463235, 10.1007/s11257-021-09294-8, marras2021, khalid2021novel, marras2019}, and augmenting explainability~\cite{10.1007/978-981-19-4453-6_4, 9439852}. While NN-based approaches have set benchmarks in predictive accuracy, this efficacy frequently comes at the cost of model interpretability, raising concerns about the trade-off between performance and transparency.

In the specific context of RL applied to KG reasoning for MOOC recommendation, existing research remains limited. To our knowledge, only two studies directly address this issue. The first, Reinforced Explainable Knowledge Concept Recommendation (EKCRec)~\cite{10.1145/3579991}, adopts an approach similar to PGPR for MOOCs, yet it has several limitations. These include the use of proprietary datasets, the absence of user studies, and the lack of publicly available code for replication. The second work presents preliminary results applying both PGPR and CAFE methodologies to publicly available MOOC datasets. However, this work remains in the discussion stage and constitutes an ongoing project~\cite{DBLP:conf/iir/AfreenBBFM23}.




In this study, we introduce an explainable MOOC recommendation system based on RL applied to KG reasoning that uses PGPR~\cite{Xian_2019}. Initially developed for product recommendation in e-commerce settings, PGPR relies on domain-specific heuristics tailored for the Amazon dataset's KG. 
In contrast, our adaptation generalizes the model to function with a new set of knowledge graphs, obviating the need for domain-specific adjustments. 
The robustness of our approach is validated through evaluations performed on two publicly accessible, real-world MOOC datasets, demonstrating its efficacy in providing both accurate and interpretable recommendations.
Additionally, in contrast to previous work, we conducted an in-depth user study to probe the alignment between end-user perception and the path-based explanations generated by our model. 
Specifically, we investigate three aspects: initially, we analyze users' preferences for path-based explanations against traditional explanations based on popularity and the behavior of similar learners (Collaborative Filtering); subsequently, we assess the alignment of path's content (teacher, course category, learner) with learners’ motivations; finally, we investigate the threshold beyond which learners begin to perceive the explanation paths as overly complex. 
The implementation is made publicly available to facilitate future research\footnote{\url{https://github.com/epfl-ml4ed/courserec}}. 

With our analyses, we aim to answer the following research questions:
\begin{enumerate}
\item[(\textbf{RQ1})] What is the performance and interpretability of path-based recommendations?
\item[(\textbf{RQ2})] What are users' preferences for explanations in terms of approach, motivation, and complexity?
\end{enumerate}

Our investigation yields several findings:
\begin{itemize}
    \item Path-based models are competitive with state-of-the-art MOOCs recommendation models in terms of accuracy.
    \item Learners showed a preference for path-based explanation compared to popularity explanation and a similar preference for the collaborative filtering explanation.
    \item Learners' motivation has a significant impact on how much detail they want in explanations; those who are learning for self-improvement want more comprehensive details.
    \item Learners do not like paths that are too long or complicated.
\end{itemize}


\section{Methodology}
\label{sec:methodology}
\begin{figure}[h]
\centering
\includegraphics[width=\textwidth]{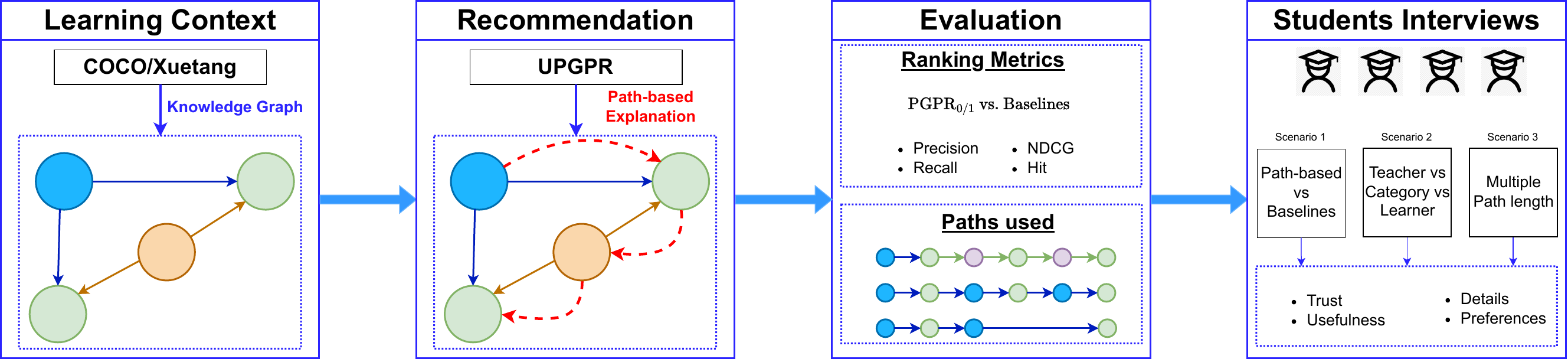}
    \caption{Explainable Recommendation and Evaluation Pipeline. We provide course recommendations for MOOCs using RL to identify optimal paths on a KG and evaluate our recommendations quantitatively and in semi-structured student interviews.}
   \label{fig:rec-eval-pipeline}
\end{figure}

With our proposed explainable recommendation approach, we aim to provide course recommendations, which are accurate and interpretable. Our pipeline for providing explainable recommendations and evaluating them is illustrated in Fig.~\ref{fig:rec-eval-pipeline}. We focus on the context of recommending courses in large MOOC platforms, using two publicly available MOOC datasets as the basis for our recommendations. In the first step (Learning Context), we define entities and relations as a basis to create a KG for each dataset. The second step (Explainable Recommendation Creation) consists of creating the explainable recommendations. Our approach, denoted as Unrestricted PGPR (UPGPR), uses RL to identify optimal paths on the KG that connect a learner to potential courses. For a learner, each recommendation is explained with the paths connecting them to the suggested course. The two final steps (Evaluation and Student Interviews) consist of a quantitative evaluation of the approach, followed by semi-structured interviews with students. 

\subsection{Learning Context}
\label{subsec:learning-context}
This step focuses on the creation of KGs to serve as input for the recommendation algorithm in the second stage. In this work, we focus on the context of course recommendations on large MOOC platforms. We therefore used two publicly available MOOCs datasets as the basis for our experiments. \texttt{COCO}~\cite{coco-dataset} and \texttt{Xuetang}~\cite{DBLP:conf/aaai/ZhangHCL0S19} are collections of online courses with enrollments, comments, as well as information on instructors and courses. Since our proposed approach is based on KGs, our first step was to build a KG for each dataset. 
In the context of MOOCs, we define the KG entities (nodes) and the KG relations (edges) as described below.  The entities are \textit{learner, school, course, category, teacher, concept}. 
\textit{Learner} denotes the learner for whom we want to recommend a \textit{course}. We defined five relations between course entities and other entities. The relation \textit{enrolled} denotes the user's enrollment in a course. The relation \textit{teaches} indicates the \textit{teacher} of a \textit{course}. The relation \textit{has\_concept} denotes that a \textit{concept} is taught in a \textit{course}. In our dataset, a course can cover multiple \textit{concepts} (e.g., python programming, project management). The relation \textit{belongs\_to} describes the \textit{category} of a \textit{course}, where each \textit{course} belongs to at least one \textit{category} (e.g., sales, STEM, media) and is taught by a \textit{teacher}. Finally, the entity \textit{school} denotes a school, and the corresponding relation \textit{provides} denotes what courses a school offers. Regarding the \texttt{COCO} dataset the \textit{concepts} taught in each course were not provided in the public dataset, so we extracted them from the course description using skillNER~\cite{DBLP:journals/eswa/FareriMCF21}. 
For both datasets, we removed learners who were enrolled in less than ten courses. The number of entities and relations of the datasets are listed in Table~\ref{table:dataset_statistics}.

\begin{table}[h]
\centering
\begin{tabular}{rrrrrr}
\toprule
\textbf{Dataset} & \textbf{Learners} & \textbf{Courses} & \textbf{Enrollments} & \textbf{KG Entities} & \textbf{KG Relations}\\ 
\midrule
COCO & 25,979 & 23,319 & 428,930 & 4,378 & 51,210 \\ 
Xuetang & 6,548 & 687 & 97,592 & 5,010 & 106,468 \\ 
\bottomrule
\end{tabular}
\caption{\textbf{Number of entities and relations in each dataset}.}
\label{table:dataset_statistics}
\end{table}


\subsection{Path-Based Explainable Recommendations Generation}
\label{subsec:pgpr-ext}
In order to provide explainable recommendations, we adapt PGPR~\cite{Xian_2019} originally used for e-commerce recommendation to our educational use case and improve it by allowing path patterns of any type and length. We call our approach Unrestricted Policy-Guided Path Reasoning (UPGPR). Below we describe the two approaches in more detail. 

\subsubsection{Policy-Guided Path Reasoning (PGPR)}
\label{subsec:model_description}
\begin{figure}[h]
\centering
\includegraphics[width=0.92\textwidth]{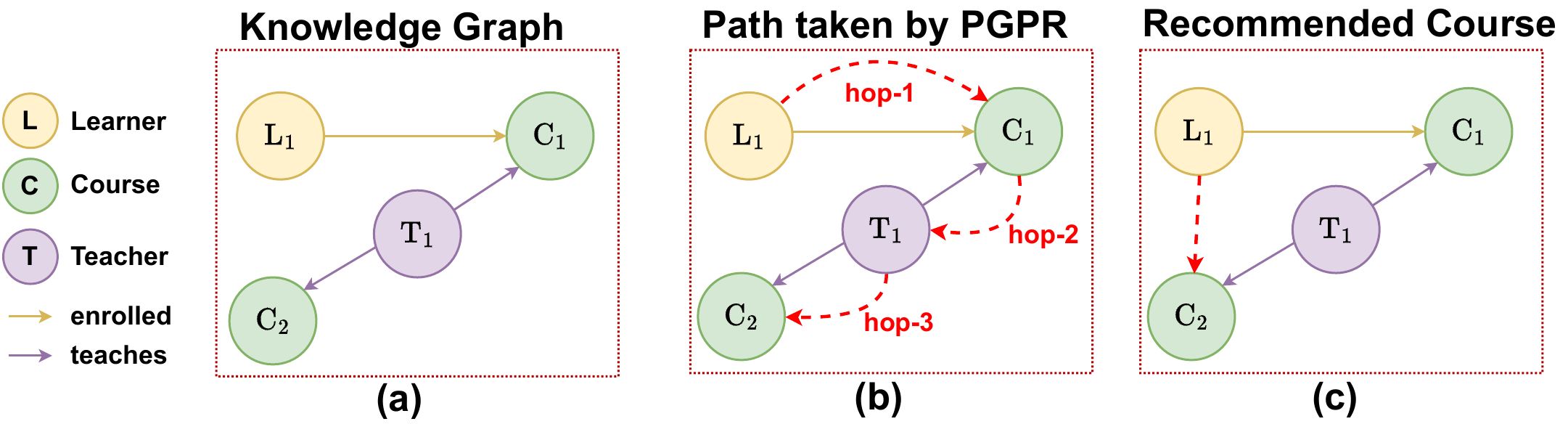}
    \caption{Illustration of the PGPR mechanism on a simple KG: \textbf{(a)} Starting from the KG, \textbf{(b)} PGPR identifies a path leading from a learner to a course. \textbf{(c)} The course reached at the end of the path is then recommended to the learner.}
   \label{fig:problem_formulation}
\end{figure}

Given a course catalog, denoted as $\mathcal{C}$, and an individual learner $l$ along with their course history, the main goal of PGPR is to pick out a suitable course from $\mathcal{C}$ to recommend to $l$. PGPR uses a graph-based approach to make these recommendations, finding paths in a knowledge graph $\mathcal{G}$ that connect the learner to a suggested course. 
A simplified example to demonstrate this is shown in Fig.~\ref{fig:problem_formulation}.

\noindent PGPR consists of two main steps: 
\begin{enumerate}[leftmargin=*]
    \item \textit{Embedding Training on KG}: first, embeddings for entities and relations found in $\mathcal{G}$ are trained. The scoring function is designed to maximize the likelihood of connecting two entities $e_i$ and $e_j$ through a relation $r$ when the triplet $(e_i,r,e_j)$ is present in the KG, and minimize it otherwise. This step allows for capturing the rich, multi-dimensional relationships among entities, thereby serving as a foundation for more accurate recommendations.
    \item \textit{RL for Pathfinding}: next, an RL agent is trained to identify optimal paths within the KG that connect a learner to potential courses. The pre-trained embeddings from the first step are integrated into the state representation of the RL agent, thereby providing enriched contextual information for decision-making. After choosing a path $p$, the agent is rewarded based on the entity reached at the end of the path. The reward $R_{\text{PGPR}}$ is defined as follows: if $p$ does not belong to a set of predefined path patterns, the agent receives a reward of $0$. Otherwise, the agent receives a reward based on the dot-product similarity between the learner $l$ embedding and the recommended course $c$ embedding. Thus, the agent is rewarded when it finds a course aligned with the learners' profile.
\end{enumerate}

\noindent A major disadvantage of the original PGPR is the use of manually predefined path patterns. While the use of these patterns allows to constrain the exploration of the agent toward paths that are relevant for explainability and end on a course, the manual definition requires time and expert knowledge. Moreover, it can be infeasible in practice for long paths where the number of different patterns grows exponentially.

\subsubsection{Unrestricted Policy-Based Path Reasoning (UPGPR)}
\label{subsec:contrib}
\begin{figure}[h]
\centering
\includegraphics[width=0.95\textwidth]{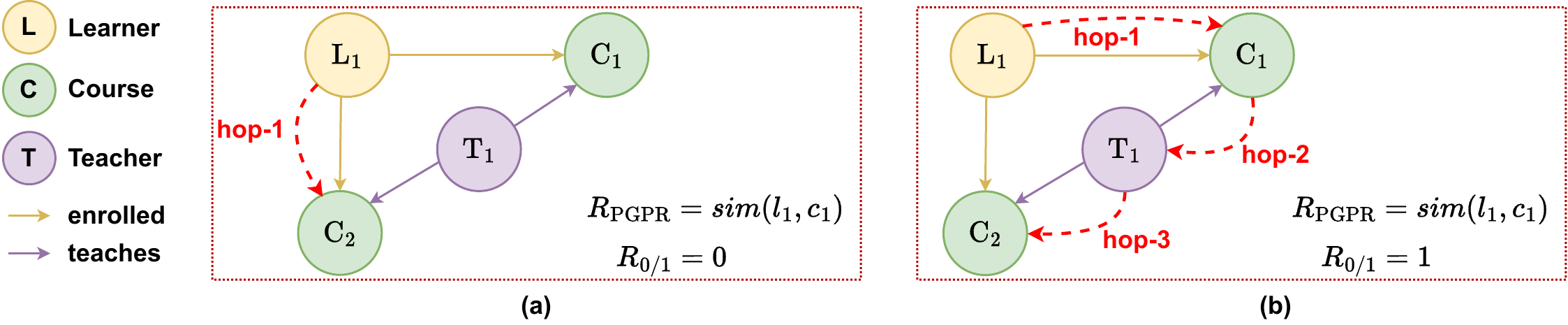}
    \caption{Illustration of the difference between traditional PGPR without path patterns and our binary reward when recommending the same course ($c_2$) to the sane learner ($l_1$) using two different paths. Both approaches incentivize the agent to find meaningful connections from a learner to a course they have enrolled in, as seen in \textbf{(b)}. However, should the agent choose a basic path leading directly from course to learner in one hop, as in \textbf{(a)}, only PGPR offers a reward.}
   \label{fig:PGPR_vs_binary}
\end{figure}

To tackle the limitations of using predefined path patterns for rewards in PGPR, we introduce a new reward that both broadens the agent's exploration and enhances generalizability, all without the need for manually set path patterns. A straightforward way to avoid path pattern limitations is to remove them from the PGPR reward equation entirely. Unfortunately, this might encourage the agent to find overly simple paths that do not generalize well. Merely removing the predefined path patterns will for example lead to the agent being rewarded for choosing the basic path shown in Fig.~\ref{fig:PGPR_vs_binary}a), leading to recommend a course the learner is already enrolled in. To solve this issue, we propose a reward that does not rely on entity or relation embeddings\footnote{The entity and relation embeddings are still being used by the agent for the state representation to help the agent model the KG.} and ignores the trivial paths that do not generalize, we propose the following binary reward:

\begin{equation}
  R_{0/1} = \begin{cases}
    1 &\text{ if } l \text{ enrolled in } c \text{ and } n_{\text{hops}}>1\\
    0,              & \text{otherwise}
\end{cases}  
\end{equation}
To be rewarded, the path taken by the agent must satisfy two conditions: it must contain more than one hop, hence avoiding rewarding trivial paths, and it must terminate on a course in which the learner (the starting entity) enrolled (see Fig.~\ref{fig:PGPR_vs_binary}b) for a comparison of the functioning of our reward versus PGPR). Our binary reward thus encourages the agent to seek out more complex paths that could potentially lead to courses not yet explored by the learner, promoting better generalization. Furthermore, the proposed reward can naturally handle path patterns of any length. In the reminder, we denote as $\text{PGPR}_{0/1}$ our model that uses the binary reward.

\subsection{Path-Based Explainable Recommendation Evaluation}
\label{subsec:quant-eval}

The goal of the evaluation is to assess the performance of the proposed recommender approach as well as the interpretability of the provided recommendations.

\noindent \textbf{Performance}. Adopting the same evaluation metrics as~\cite{Xian_2019}, we employed Normalized Discounted Cumulative Gain (NDCG), Recall, Hit Ratio (HR), and Precision.
We compared our approach and PGPR against three baselines: Pop, NeuMF~\cite{DBLP:conf/www/HeLZNHC17}, and CFKG~\cite{DBLP:journals/algorithms/AiACZ18}.
Pop ranks courses based on their popularity, defined as the frequency of recommendation in the training set. Pop is not a personalized recommendation approach since it does not take learner preference into account when recommending. NeuMF is a state-of-the-art collaborative filtering approach that jointly utilizes matrix factorization and a multi-layer perceptron to generate a matching score between a learner and course. CFKG is a state-of-the-art approach that uses neural networks and KG.

\noindent \textbf{Interpretability}. Since we do not employ predefined patterns, there is a possibility that the paths utilized by the agents may lack interpretability. This is especially true for longer paths. To analyze the interpretability of our model, we present the most frequent paths used by the agent on the test set. We excluded all the non-valid paths, i.e. the ones that do not end on a course.

\subsection{Qualitative User Study}
\label{subsec:user-study}
We conducted semi-structured interviews with 32 participants to evaluate learners' preferences regarding explanations, the alignment of explanations with their objectives, and the complexity of the path-based explanations. Of these participants, 7 contributed to a pilot study, assisting us in developing and refining motivational scenarios and items. The final study was then carried out with the remaining 25 participants.

\noindent\textbf{Participants}. We recruited the $25$ Ph.D. students and postdocs mainly through email and posts in Slack channels. Participants’ ages spanned from $21$ to $50$ years, with an average age of ($\mu=29.6, \sigma=6.8$). The majority of participants ($76\%$) identified as male and the remaining ones as female. Participants originated from $11$ distinct countries. All participants possessed a background in STEM (Science, Technology, Engineering, and Mathematics) disciplines. We obtained informed consent to participate in the study and to record their interviews.

\noindent\textbf{Procedure}. We conducted semi-structured interviews using a course recommendation context. Participants were asked to assume the role of a Ph.D. student choosing a course on a MOOC platform. 
To assess the influence of learners’ motivations on their explanation preferences, participants were randomly allocated to one of two conditions, each representing different learning motivations: 1) \textit{Learn}, instructing participants to assume that their primary goal was to develop a skill pertinent to their Ph.D. research or 2) \textit{Credits}, instructing participants to assume that their primary aim was to fulfill their Ph.D. credit requirements. The objective was to investigate whether intrinsic motivations (e.g. learning skills) and extrinsic incentives (e.g. credit fulfillment) affected explanation preferences in a manner analogous to their impact on performance~\cite{cerasoli2014intrinsic}. All other aspects of the interview remained the same for all participants.

\begin{figure}[t]
    \centering
    \includegraphics[trim={0 3.1cm 0 5.8cm}, clip, width=0.95\textwidth]{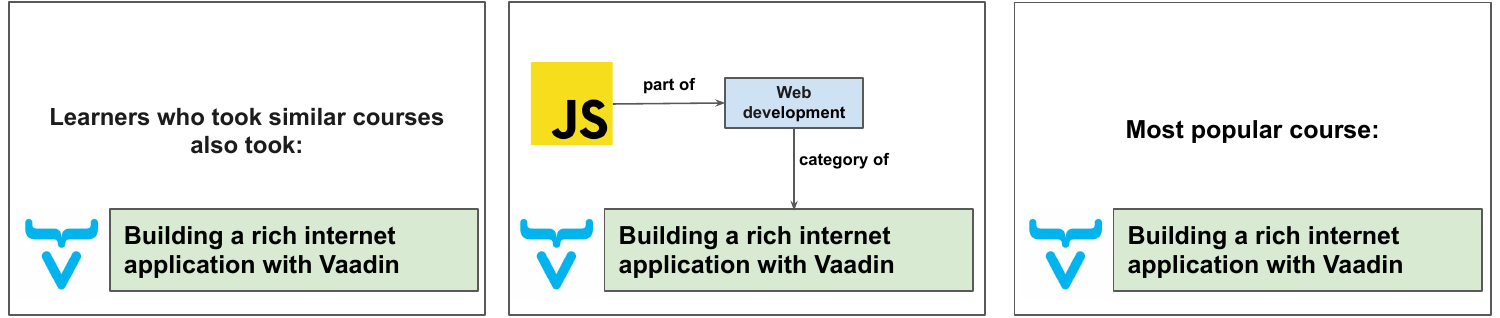}
    \caption{\textit{1 - Algorithm}: Comparison of collaborative filtering (left), path-based (middle), and popularity-based (right) explanations.}
    \label{fig:scenario_1}
    \vspace{3mm}   
    \includegraphics[trim={0 3.5cm 0 6.1cm}, clip, width=0.95\textwidth]{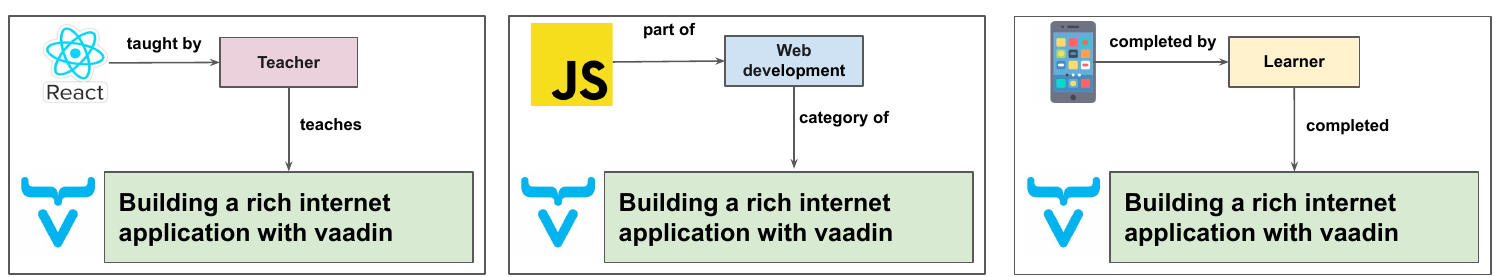}
    \caption{\textit{2 - Motivation}: Comparison of different relations, one of them being aligned to the learner's motivation (skill development or credit fulfillment): Teacher-relation (left), Category-relation (middle), and learner relation (right).}
    \label{fig:scenario_2}
    \vspace{3mm}   
    \includegraphics[trim={0 2.5cm 0 5.65cm}, clip, width=0.95\textwidth]{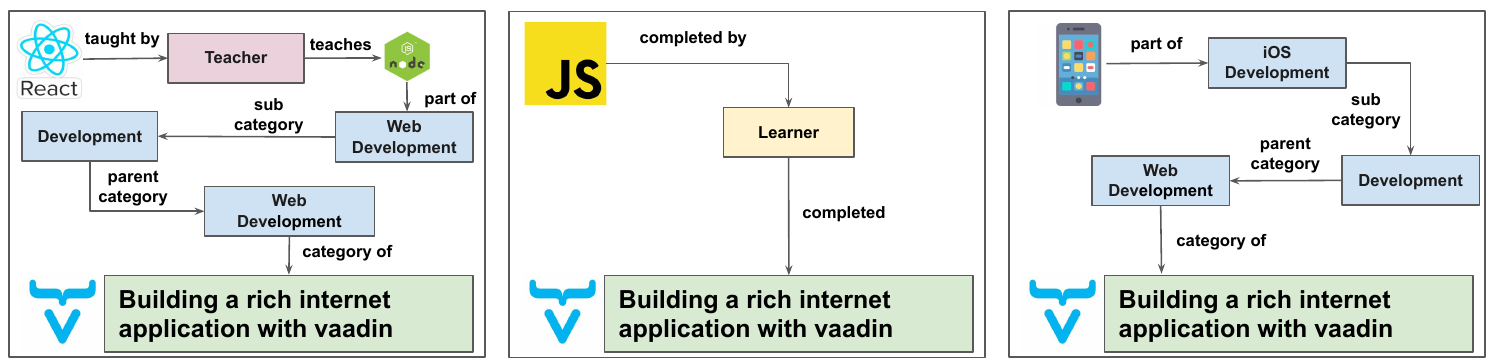}
    \caption{\textit{3 - Complexity}: Comparison of different path lengths: length $6$ (left), length $2$ (middle), and length $4$ (right).}
    \label{fig:scenario_3}
\end{figure}

To emulate a learner experience within a MOOC platform, participants were then instructed to consider themselves as having completed three courses from a given list. A recommendation system then suggested a course to the participants. To avoid participants being biased by a preference for some of the completed courses, we assumed a high rating (in terms of stars) for each course. Participants were then presented with three distinct scenarios (see Figs.~\ref{fig:scenario_1},~\ref{fig:scenario_2}, and~\ref{fig:scenario_3}), each featuring three different explanations for the recommended course:
\begin{enumerate}[leftmargin=*]
    \item \textit{1 - Algorithm}: path-based explanation that uses the category of the courses, explanation based on collaborative filtering, and popularity-based explanation (Fig.~\ref{fig:scenario_1}). The goal of this scenario was to study learner preferences between a path-based explanation and conventional explanations.
    \item \textit{2 - Motivation}: path-based explanations using three different relations (Fig.~\ref{fig:scenario_2}), one of them being aligned to the learners' motivation. The goal of this scenario was to study learner preferences related to their purported motivation.
    \item \textit{2 - Complexity}: path-based explanations using a path of length $2$ (equivalent to using 2 relations in the path), $4$, and $6$ (Fig.~\ref{fig:scenario_3}). The goal of this scenario was to study the clarity of paths of different lengths.
\end{enumerate}

\noindent For each scenario, we first asked participants which explanation they preferred and why. Then, participants were given four items from the Explanation Satisfaction Scale~\cite{DBLP:journals/corr/abs-1812-04608} (we selected the subset of items relevant to our context):
\begin{enumerate}
    \item[Q1] \textit{"This explanation of how the recommendation system works is useful to my goals."}
    \item[Q2] \textit{"This explanation lets me judge when I should trust the recommendation system."}
    \item[Q3] \textit{"This explanation of how the recommendation system works has sufficient detail."}
    \item[Q4] \textit{"This explanation of how the recommendation system works has irrelevant detail."}
\end{enumerate}

\noindent Participants were asked to indicate their agreement with each of the statements using a 5-point Likert scale ranging from "strongly disagree" to "strongly agree".

\noindent\textbf{Statistical analysis}. For our statistical analysis of the Likert scale responses, we first examined normality with the Shapiro-Wilk test and determined that ANOVA assumptions weren't met. Hence, we applied the non-parametric Kruskal-Wallis test to discern significant variations among responses within each scenario. For post-hoc pairwise analyses, we compared responses for each explanation and between the \textit{Learn} and \textit{Credits} conditions using the Mann-Whitney U Test and adjusted for multiple comparisons using the Benjamini-Hochberg method. Throughout the rest of this paper, any mention of a statistically significant difference implies $p<0.05$ for the related test.








\section{Results}
\label{sec:results}
We conducted experiments on the \texttt{COCO} and \texttt{Xuetang} datasets to assess the performance of our approach compared to baseline approaches and to study the impact of different path lengths on interpretability and performance (RQ1). We then evaluated the semi-structured interviews to assess students' preferences in terms of recommendation algorithm (path-based vs. conventional), alignment to motivation (skill development vs. credit fulfillment), and path length (RQ2). 

\subsection{Experimental Protocol}
\label{subsec:exp-protocol}
To evaluate the performance of our approach, we split the \texttt{COCO} and \texttt{Xuetang} user-item interaction data into train, test, and validation sets in a ratio of $80\%$:$10\%$:$10\%$ for each user. We trained and evaluated each model (our approach and the baselines) three times with different seeds.
We implemented and evaluated the baselines using Recbole~\cite{DBLP:conf/cikm/ZhaoMHLCPLLWTMF21}. We used batches of size $512$ and the Adam optimizer with a learning rate of $0.001$. We trained the models for $200$ epochs with early stopping on the validation set with a patience of $10$ epochs. In developing UPGPR, we utilized the same hyperparameters (batch size, learning rate, and embedding size) for training KG embeddings as in the original PGPR implementation~\cite{Xian_2019}. The RL agents were trained for 50 epochs, using Adam optimizer with a learning rate of 0.001. During the training phase, to encourage the agent's exploration, we permitted the agent to traverse a path that could extend up to +1 longer than permitted during the evaluation phase.

\subsection{Performance and Interpretability of Recommender (RQ1)}
\label{subsec:quantanalyses-results}
In our first analysis, we evaluated the performance of our approach in comparison to popular baselines. Furthermore, we also assessed the influence of the path length on performance and interpreted the recommended paths.

\vspace{1mm} \noindent \textbf{Performance.} Table~\ref{tab:experiment-1-table} details the performances of our approach as well as all baseline models. The best results are highlighted in bold. We observe that our proposed model surpasses all the baseline models across all ranking metrics. The popularity-based model (Pop) is the least proficient, mainly because of its lack of personalization since every learner receives the same recommendations regardless of their history. Notably, our approach also outperforms state-of-the-art collaborative filtering (NeuMF) and neural-network-based models (CFKG), indicating that the proposed approach provides explainable recommendations without compromising accuracy.

In instances where the path length is limited to $3$, our model displays similar performance to the original PGPR model. However, PGPR requires expert knowledge to provide manually defined path patterns (valid paths). With an increasing path length, this task becomes exponentially complex due to the exponential increase in the number of feasible paths. In contrast, our approach does not require predefined path patterns and can handle paths of any length, enabling an automatic transfer to different data sets and recommendation scenarios.

We observe that for our approach, a longer path is typically associated with better performance. This implies that, even with the exponential increase of potential paths, the agent can find more pertinent paths, leading to better recommendations. However, we hypothesized that longer paths are more difficult to interpret (see also the user study results in Section~\ref{subsec: userstudy-results}) and therefore calculated results only up to a path length of $5$.

\begin{table}[h]
  \centering
\scalebox{0.8}{
  \begin{tabular}{lllcllllll}
\toprule
\textbf{Dataset} & \textbf{Model} & \textbf{Type} & \textbf{Path Length} & \textbf{NCDG} & \textbf{Recall} & \textbf{HR} & \textbf{Precision} & \textbf{Invalid users}\\
\midrule
\multirow{8}{*}{\texttt{Xuetang}} & Pop & Popularity & - & 03.06 \textpm\ 0.0 & 04.80 \textpm\ 0.0 & 09.30 \textpm\ 0.0 & 00.97 \textpm\ 0.0 & 00.0\% \textpm\ 0.0 \\
& NeuMF & Collaborative Filtering & - & 14.05 \textpm\ 0.6 & 24.15 \textpm\ 1.2 & 39.01 \textpm\ 1.7 & 04.39 \textpm\ 0.2 & 00.0\% \textpm\ 0.0 \\
& CFKG & Neural Network  & - & 14.20 \textpm\ 0.9 & 24.77 \textpm\ 1.5 & 39.88 \textpm\ 2.1 & 04.52 \textpm\ 0.3 & 00.0\% \textpm\ 0.0 \\
& PGPR & Path-Based & 3 & 18.16 \textpm\ 0.1 & 24.72 \textpm\ 0.1 & 41.03 \textpm\ 0.3 & 04.75 \textpm\ 0.0 & 04.4\% \textpm\ 0.5\\
\cmidrule{2-9}
& UPGPR & Path-Based  & 3 & 18.43 \textpm\ 0.0 & 25.03 \textpm\ 0.2 & 41.44 \textpm\ 0.3 & 04.26 \textpm\ 0.0 & 04.8\% \textpm\ 0.2 \\
& UPGPR & Path-Based  & 4 & 17.96 \textpm\ 0.1 & 24.47 \textpm\ 0.1 & 39.8 \textpm\ 0.2 & 04.63 \textpm\ 0.0 & 06.8\% \textpm\ 1.0\\
& UPGPR & Path-Based  & 5 & \textbf{20.85} \textpm\ 0.1 & \textbf{30.38} \textpm\ 0.2 & \textbf{47.78} \textpm\ 0.4 & \textbf{05.64} \textpm\ 0.1 & 00.0\% \textpm\ 0.0 \\
\midrule
\multirow{8}{*}{\texttt{COCO}} & Pop & Popularity & - & 3.90 \textpm\ 0.0 & 6.21 \textpm\ 0.0 & 10.66 \textpm\ 0.0 & 1.14 \textpm\ 0.0 & 00.0\% \textpm\ 0.0 \\
& NeuMF & Collaborative Filtering & - & 5.55 \textpm\ 0.1 & 9.40 \textpm\ 0.3 & 16.45 \textpm\ 0.5 & 1.77 \textpm\ 0.1 & 00.0\% \textpm\ 0.0 \\
& CFKG & Neural Network  & - & 7.79 \textpm\ 0.0 & 13.06 \textpm\ 0.0 & 22.23 \textpm\ 0.1 & 2.43 \textpm\ 0.0 & 00.0\% \textpm\ 0.0 \\
& PGPR & Path-Based  & 3 & 07.47 \textpm\ 0.2 & 10.35 \textpm\ 0.2 & 18.7 \textpm\ 0.3 & 02.08 \textpm\ 0.0 & 00.3\% \textpm\ 0.1\\
\cmidrule{2-9}
& UPGPR & Path-Based  & 3 & 06.66 \textpm\ 0.1 & 09.19 \textpm\ 0.1 & 16.55 \textpm\ 0.2 & 01.82 \textpm\ 0.0 & 00.5\% \textpm\ 0.0 \\
& UPGPR & Path-Based  & 4 & 08.49 \textpm\ 0.1 & 12.35 \textpm\ 0.1 & 21.73 \textpm\ 0.0 & 02.42 \textpm\ 0.0 & 00.2\% \textpm\ 0.0\\
& UPGPR & Path-Based  & 5 & \textbf{08.87} \textpm\ 0.1 & \textbf{13.64} \textpm\ 0.1 & \textbf{23.77} \textpm\ 0.3 & \textbf{02.64} \textpm\ 0.0 & 00.0\% \textpm\ 0.0\\
\bottomrule
\end{tabular}
 }
  \caption{Performance comparison (mean\textpm standard deviation) across the two recommendation datasets of our new reward with different path lengths against PGPR and popular baselines. For `Invalid users', the agents could not recommend at least $10$ items. The highest metric for each dataset is highlighted in bold.}
  \label{tab:experiment-1-table}
\end{table}

\begin{figure}[h]
\centering
\includegraphics[width=\textwidth]{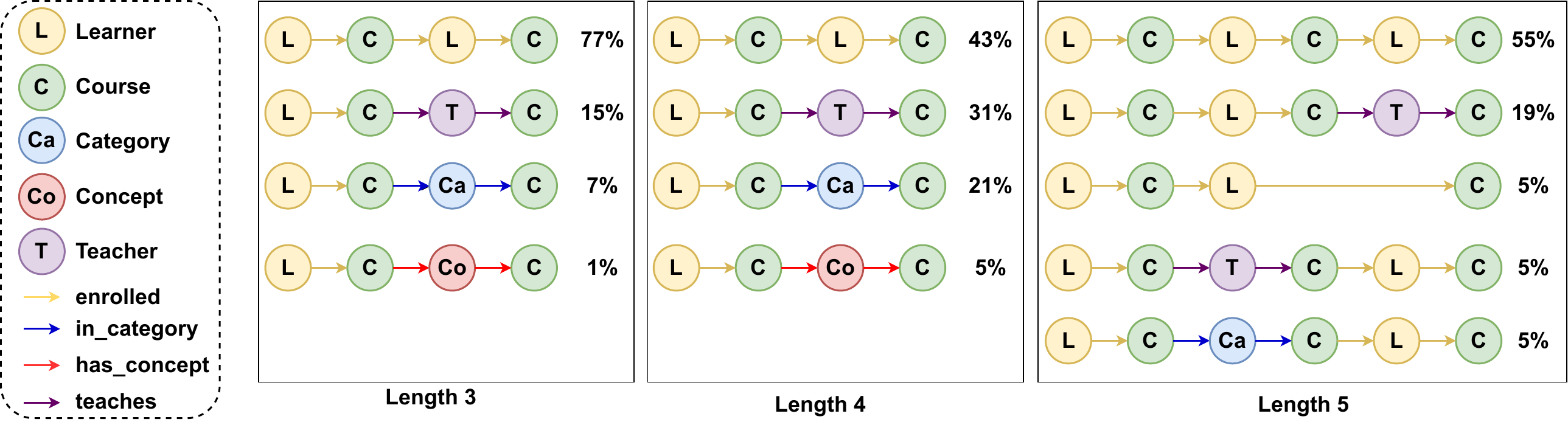}
    \caption{Path patterns discovered by our agents when trained with different path lengths using UPGPR on the \texttt{COCO} dataset. For paths of length 5, we only display the 5 most frequent patterns.}
   \label{fig:pattern_interpretation}
\end{figure}

\noindent \textbf{Interpretability.} Fig.~\ref{fig:pattern_interpretation} displays the path patterns used by our agents during the evaluation phase on the \texttt{COCO} dataset. For a path length of $3$ ($3$ hops), there exist only four different valid patterns. This restriction is due to the structure of the KG of the dataset: a \textit{user} is connected only to a \textit{course} (through the \textit{enrolled} relation), and there are now connections between for example users a and categories. We observe that the predominant pattern based on shared course enrollment between users is selected by the agent $77\%$ of the time. This pattern is similar to user-based collaborative filtering. The second most frequent pattern, chosen in $15\%$ of the cases, recommends to users a course from an already known teacher. The two last patterns are recommendations based on the course content: the user is recommended a course from the same category ($7\%$) or teaching a similar concept ($1\%$) as an already visited course.

For paths of length of $4$, the agent uses the exact same patterns as for a path length of $3$. this result is again due to the structure of the underlying KG: since it is not possible to go from a user to a course in $4$ hops, the learned policy includes paths with $3$ hops. However, extending the path length leads to a more balanced distribution of the observed patterns. While the agent still uses shared course enrollments between users the most ($43\%$), the teacher-based ($31\%$) and category-based patterns ($31\%$) are also selected frequently.

For paths of length $5$, the agent adopts a more varied approach, utilizing $18$ different patterns. Fig.~\ref{fig:pattern_interpretation} illustrates the five predominant patterns, which collectively represent $89\%$ of the paths employed by the agent. Four of the most frequent patterns are of length $5$. The top pattern is again a path going through learners and courses, selected in $55\%$ of the cases. Two other frequent patterns use a combination of relations between courses and teachers and courses and learners. In one case, a first course is reached via shared enrollment between users and the final recommendation is a course taught by the same teacher as the first course ($19\%$). The other case uses the opposite patterns, reaching a first course via the teacher and the final recommendation via shared enrollment ($5\%$). One pattern involves category: a first course is reached via the category relationship and the second path again via shared enrollment ($5\%$). Finally, there is one path of length $3$, which corresponds to the shared enrollment of users pattern ($5\%$) found for path lengths of $3$ and $4$.

Interestingly, paths incorporating concept and category entities are infrequently used by the agents. This is likely due to the limited number of relations involving concepts and categories available in the KG. To quantify, while a substantial portion of the relations in the \texttt{COCO} KG is represented by learners enrolled in courses ($89.3\%$) and teachers teaching courses ($3.8\%$), the relations involving categories and concepts comprise $4.8\%$ and $2.1\%$of the KG, respectively.

\subsection{Users' Explanation Preferences (RQ2)}
\label{subsec: userstudy-results}
In our second analysis, we evaluated the data collected in our $25$ semi-structured interviews to analyze the alignment of the provided explanations with users' preferences and motivations. 

\noindent \textbf{Algorithm Preference.} In scenario \textit{1 - Algorithm}, $13$ participants favored the path-based explanation, $10$ opted for the collaborative filtering, and only one for the popularity-based. Another participant was undecided. Overall, there was a mild preference for the path-based over the collaborative filtering explanation. Regarding the less-favored popularity approach, a participant commented: “It’s not specific for me, it might not fit my needs”.

To understand the participants' preferences, we analyzed their open-ended responses.
The primary reasons we identified were:
"Understanding and agreeing with the explanation" (mentioned by 11 participants);
"Detailed and informative" (noted by 10 participants);
"Affective judgment" (highlighted by 4 participants).
One participant commented on the path-based explanation, stating: "there is a logical reason for the recommendation." In contrast, another participant found the collaborative filtering explanation to be "more personal." They added that the other two explanations were "not personal at all." This suggests that learners convinced by formal reasoning for the recommendation might lean towards path-based approaches. On the other hand, those seeking more personalized content might favor collaborative filtering, which is based on similar learners.

\begin{figure}[t]
    \centering
    \includegraphics[width=\textwidth]{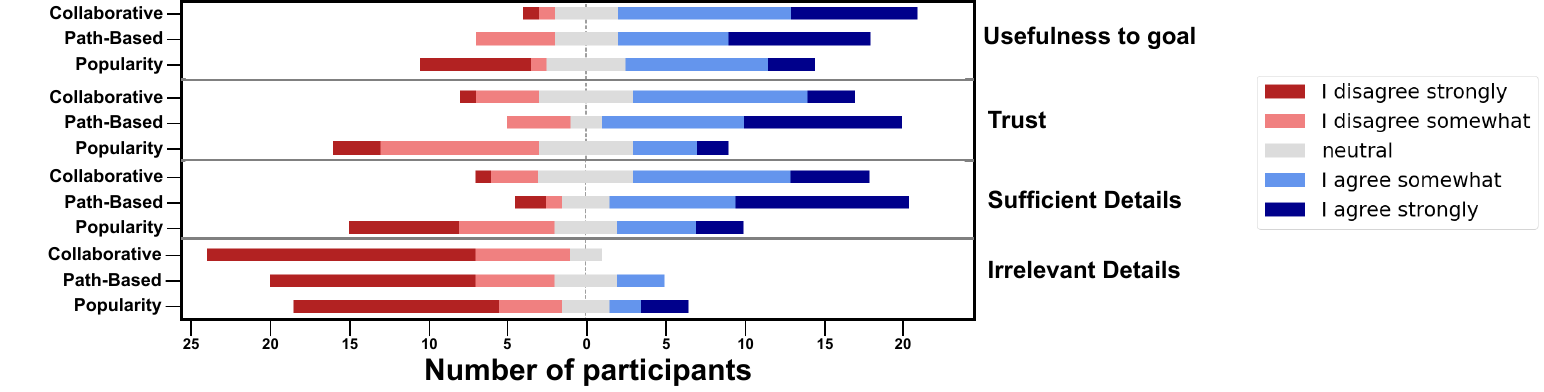}
    \caption{Participants' answers for the Likert-scale items of scenario \textit{1 - Algorithm}.}
    \label{fig:likert_scenario_1}
\end{figure}

In addition, we evaluated participants' answers to the Likert-scale items. Given the nature of the path-based explanations, we expected to find the differences between the algorithms predominantly in the \textit{Sufficient Details} item (as path-based recommendations provide more detailed explanations) and the \textit{Trust} item (assuming that the more detailed path-based explanation would increase participants' trust~\cite{DBLP:journals/bjet/NazaretskyACA22}). We did not expect to see differences between the two conditions. Participants' responses can be seen in Fig.~\ref{fig:likert_scenario_1}, with averages and standard deviations detailed in Table~\ref{tab:diff}.

Indeed, participants' trust scores varied notably across algorithms. The path-based explanation was rated the highest in trust ($\mu=4.00$), followed by collaborative filtering ($\mu=3.44$) and popularity ($\mu=2.68$).
The low trust in the popularity explanation was evident, with 9 participants criticizing its use for recommendations. One participant mentioned that this approach was "biased due to learners who took them". This confirms our previous findings that popularity is usually not a good standalone argument when recommending courses.

Regarding the \textit{Sufficient Details}, the path-based explanation is perceived as the most detailed ($\mu=4.00$), followed by collaborative filtering ($\mu=3.60$), and popularity ($\mu=2.64$). We found significant differences between the algorithms. Pairwise comparisons also showed significant differences between  Collaborative Filtering and Popularity  as well as between Path-based and Popularity. Participants in the \textit{Learn} condition valued the detail in the path-based explanations more than those in the \textit{Credit} condition($\mu_{Learn}=3.50$, $\mu_{Credits}=4.45$), indicating a greater detail demand for such learners despite the fact that detail was rarely cited as the main reason for preference.

As expected, the participants generally did not perceive the provided explanations as containing irrelevant details.

Regarding the \textit{Usefulness to Goal}, both the path-based ($\mu=3.80$) and the collaborative filtering ($\mu=3.96$) explanations were perceived as useful. Based on these findings, it seems that the usefulness of the explanations was not the main driver for the choice of the preferred algorithm. This is in alignment with the open-ended answers, where only a single participant mentioned the goal when asked about the reason for their preferences.

\vspace{1mm} \noindent \textbf{Alignment with Motivation.} For the second scenario (\textit{2 - Motivation}), we anticipated \textit{Learn} participants would favor the path-based explanation using category due to their skill development motivation. 
For the \textit{Credits} condition, we hypothesized a familiar teacher would be appealing. Results showed a preference for paths using category in both conditions with $12$ participants preferring Category, $8$ preferring Teacher, and $3$ preferring Learner. Six participants critiqued the insufficient info about learners. 

Open-ended answers primarily indicated two reasons for these preferences:
They were "relevant/helpful/useful" (mentioned by 10 participants).
They provided clarity (highlighted by 6 participants).
Participants' views varied. If they believed the teacher significantly influenced the course, they leaned toward teacher-based explanations. One participant commented, "Teacher significantly affects the quality of the course." Another said, "A teacher in common would be more trusted than a learner in common."
If they were more focused on the course outcome or subject, they favored the category. For instance, one participant noted the explanation was "not helpful; it doesn’t specify the course output or the prerequisite knowledge", Another comment was, "In MOOCs, the teacher matters, but the subject is paramount."
A different participant stressed, "Prioritize the subject over the teacher or other learners." Meanwhile, another found the category preferable because it "provides direct information about the topic." This suggests that the preferences in explanations reflect the aspect of the course the learner deems the most important.

\begin{figure}[t]
    \centering
    \includegraphics[width=\textwidth]{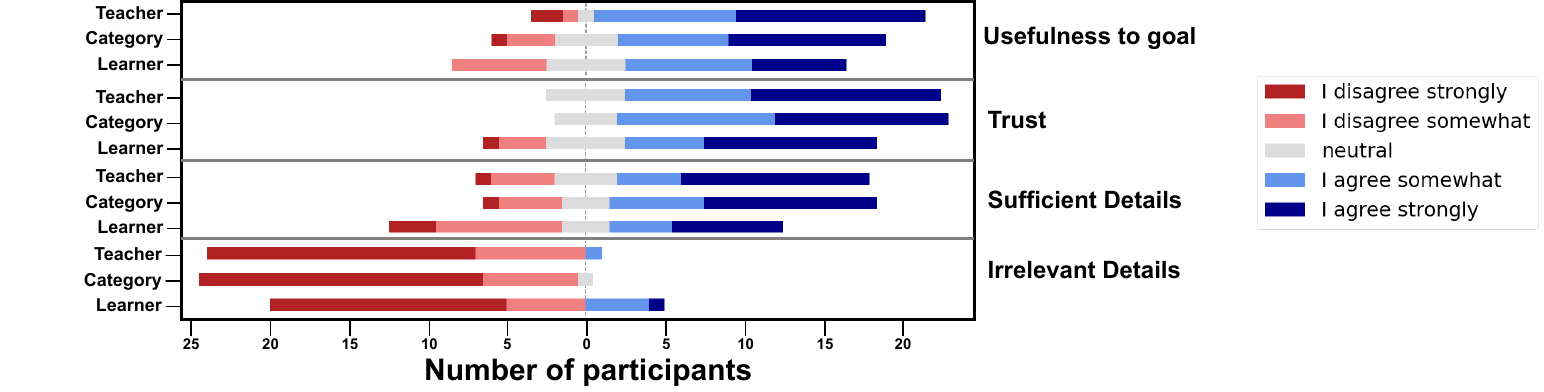}
    \caption{Participants' answers for the Likert-scale items of scenario \textit{2 - Motivation}.}
    \label{fig:likert_scenario_2}
\end{figure}

Again, we evaluated participants' responses to the Likert-scale items (see Fig. ~\ref{fig:likert_scenario_2}). Due to the design of the scenario, we only expected to find differences in scores for the \textit{Usefulness to Goals} item related to the condition. 

As expected, the \textit{Credits} group, the teacher path was deemed most useful ($\mu=4.31$), followed by category ($\mu=3.85$) and learner ($\mu=3.62$). In contrast, \textit{Learn} participants showed similar scores for category and teacher (both $\mu=3.92$), with a lesser score for learner ($\mu=3.50$). The usefulness seemed constrained by the explanations' information volume. Five \textit{Learn} participants pointed out insufficient course information, an issue not raised by the \textit{Credits} group, highlighting the \textit{Learn} group's need for more detailed data.

Confirming our hypothesis, participants demonstrated a high trust in explanations using category and teacher, with one participant mentioning that they were both "fully transparent". While the majority of participants also perceived the path through the learner as trustworthy ($17$ participants agreed or strongly agreed), the average score of this path was a bit lower ($\mu_{teacher}=4.28$, $\mu_{category}=4.28$, $\mu_{learner}=3.88$). As stated previously, participants want to know more about the learner with a participant stating that they would "need to know a bit more about learner".

The lack of information about the learner was evident in the \textit{Sufficient Details} responses. For the path going through the learner, $11$ participants perceived that they were not provided with sufficient details (``disagree somewhat or strongly'') with one participant mentioning that "details about the type of learner are missing". Interestingly, we again found significant differences between the two conditions: participants in the \textit{Credits} condition were more satisfied with the degree of detail provided by the teacher and category explanations than participants in the \textit{Learn} condition.


Finally, the majority of participants identified few or no irrelevant details within the path-based explanations, attributing this to the overall succinctness and relevance of the information provided in the explanations. Again, we observe a slightly higher average score for learner ($\mu=1.85$) than for teacher ($\mu=1.41$) and category ($\mu=1.33$)

\vspace{1mm} \noindent \textbf{Influence of path length.} In the \textit{3 - Complexity} scenario, we examined path lengths and hypothesized that longer paths could be challenging to interpret due to the high amount of details. As expected, the longest path was the least favored: $11$ participants preferred the path of length $4$ (medium), $10$ participants preferred the path of length $2$ (shortest), and only two participants preferred the path of length $6$ (longest). Two participants showed no preference.

To understand participants' preferences, we delved into their open responses. Two main themes arose:
Complexity/confusion (cited by 12 participants).
Detail/information/interpretability (brought up by 8 participants).
A participant, referencing the longest path, commented, "Why should I read all the details? Too much detail." Another observed that the longest explanation "provides the most detailed information, but it's not user-relevant." These responses emphasize the need for explanations that are relevant and not overly detailed.
Another key point, stressed by 4 participants, was the necessity for a logical and specific connection between learners and courses. Feedback on the longest explanation included comments like "steps between the entities seem arbitrary",  "the connection between the courses I took and the ones recommended isn't clear.", and  "steps between the entities could connect any two entities". This feedback indicates that a long path or a path perceived as illogical will be deemed irrelevant or non-informative.
However, feedback on the medium explanation was more favorable, with remarks like "useful because the connections make sense." This implies that participants can appreciate a medium-length path if the relationship between previously taken and recommended courses is comprehensible.

\begin{figure}[h]
    \centering
    \includegraphics[width=\textwidth]{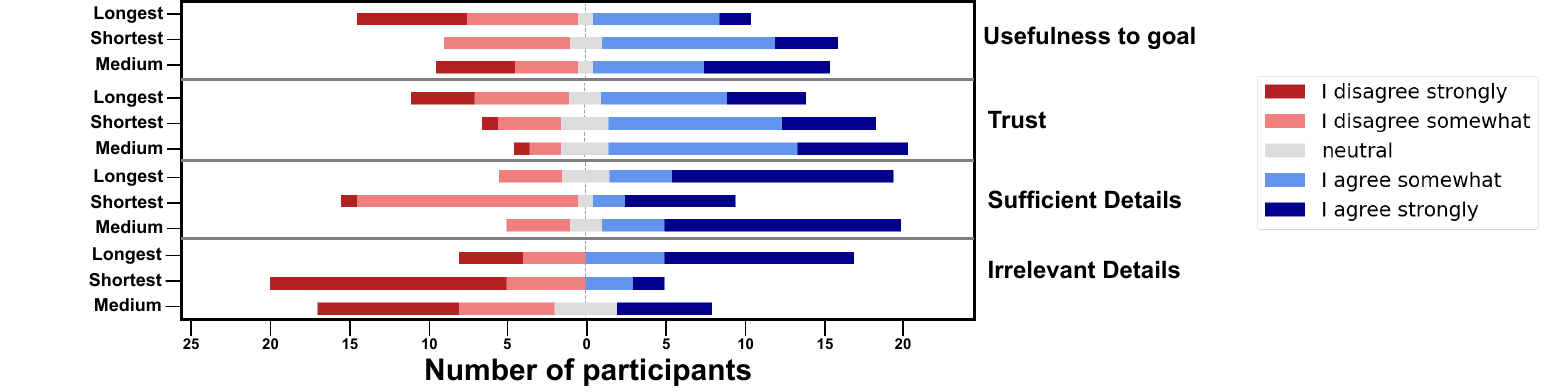}
    \caption{Participants' answers for the Likert-scale items of scenario \textit{3 - Complexity}}
    \label{fig:likert_scenario_3}
\end{figure}

For the \textit{3 - Complexity} scenario, we expected differences in the \textit{Irrelevant Details} item based on path lengths and lower \textit{Usefulness} scores for longer paths, anticipating comprehension issues. We didn't foresee any condition differences. Participants’ responses are in Fig.~\ref{fig:likert_scenario_3}.

Longer paths were perceived as having more irrelevant details, emphasizing a preference for concise explanations with a participant stating that it "is way too crowded". $17$ participants perceived the longest path as containing irrelevant details (``agree strongly or somewhat''). In contrast, for the medium and the shortest path, this was the case for only $8$ and $5$ participants respectively. Pairwise comparisons found significant differences between the longest and shortest path ($p = 0.001$) and the longest and medium path ($p=0.043$). These findings are in alignment with the fact that complexity was one of the main topics mentioned in the open-ended answers.

As anticipated for the \textit{Sufficient Details} item, $19$ participants found the medium path, $18$ found the longest path adequately detailed, and only $9$ found the shortest path sufficiently detailed ($\mu_{Shortest}=3.0$). This contrasts with a previous scenario where paths of similar length were seen as detailed ($\mu_{1-Algorithm}=4.0$). The difference might stem from the paths used for comparison (users might have realized that longer paths can provide even more information) and the fact that the shortest path in this scenario went through the learner, which was consistently rated lower.




\begin{table}[h]
\centering
\resizebox{\textwidth}{!}{
\begin{tabular}{llcccccccccccc}
\toprule
\multirow{2}{*}{\textbf{Scenario}}& \multirow{2}{*}{\textbf{Explanation}} & \multicolumn{3}{c}{\textbf{Usefulness to Goals}} & \multicolumn{3}{c}{\textbf{Trust}} & \multicolumn{3}{c}{\textbf{Sufficient details}} & \multicolumn{3}{c}{\textbf{Irrelevant details}} \\
\cmidrule(lr){3-5}\cmidrule(lr){6-8}\cmidrule(lr){9-11}\cmidrule(lr){12-14}
& & \textit{All} & \textit{Learn} & \textit{Credits} & \textit{All} & \textit{Learn} & \textit{Credits} & \textit{All} & \textit{Learn} & \textit{Credits} & \textit{All} & \textit{Learn} & \textit{Credits} \\
\midrule
\multirow{3}{*}{\textit{1 - Algorithm}} & Collaborative & $3.96$ ($1.0$) & $3.58$ ($1.2$) & $4.31$ ($0.8$) & $3.44$ ($1.0$) & $3.33$ ($1.1$) & $3.54$ ($1.1$) & $3.60$ ($1.1$) & $3.58$ ($0.9$) & $3.62$ ($1.3$) & $1.40$ ($0.6$) & $1.58$ ($0.7$) & $1.23$ ($0.6$)\\
& Path-based & $3.80$ ($1.2$) & $3.33$ ($1.2$) & $4.23$ ($1.0$) & $4.00$ ($1.1$) & $3.92$ ($1.1$) & $4.08$ ($1.1$) & $4.00$ ($1.2$) & $3.50$ ($1.5$) & $4.46$ ($0.7$) & $1.88$ ($1.1$) & $2.08$ ($1.0$) & $1.69$ ($1.2$)\\
& Popularity & $3.00$ ($1.4$) & $2.42$ ($1.4$) & $3.54$ ($1.3$) & $2.68$ ($1.1$) & $2.58$ ($1.1$) & $2.77$ ($1.2$) & $2.64$ ($1.4$) & $2.92$ ($1.5$) & $2.38$ ($1.3$) & $2.12$ ($1.5$) & $2.33$ ($1.6$) & $1.92$ ($1.4$)\\
\hline
\multirow{3}{*}{\textit{2 - Motivation}} & Teacher & $4.12$ ($1.2$) & $3.92$ ($1.1$) & $4.31$ ($1.3$) & $4.28$ ($0.8$) & $4.08$ ($0.8$) & $4.46$ ($0.8$) & $3.88$ ($1.3$) & $3.17$ ($1.4$) & $4.54$ ($0.8$) & $1.40$ ($0.7$) & $1.58$ ($0.9$) & $1.23$ ($0.4$)\\
& Category & $3.88$ ($1.2$) & $3.92$ ($1.2$) & $3.85$ ($1.3$) & $4.28$ ($0.7$) & $4.08$ ($0.7$) & $4.46$ ($0.8$) & $3.88$ ($1.3$) & $3.33$ ($1.4$) & $4.38$ ($1.0$) & $1.32$ ($0.6$) & $1.42$ ($0.7$) & $1.23$ ($0.4$)\\
& Learner & $3.56$ ($1.1$) & $3.50$ ($1.1$) & $3.62$ ($1.2$) & $3.88$ ($1.2$) & $3.58$ ($1.2$) & $4.15$ ($1.3$) & $3.16$ ($1.5$) & $3.00$ ($1.3$) & $3.31$ ($1.7$) & $1.84$ ($1.3$) & $2.00$ ($1.3$) & $1.69$ ($1.3$)\\
\hline
\multirow{3}{*}{\textit{3 - Complexity}} & Length-6 & $2.64$ ($1.4$) & $2.92$ ($1.4$) & $2.38$ ($1.4$) & $3.16$ ($1.4$) & $3.25$ ($1.3$) & $3.08$ ($1.6$) & $4.12$ ($1.2$) & $3.58$ ($1.4$) & $4.62$ ($0.7$) & $3.68$ ($1.6$) & $3.50$ ($1.7$) & $3.85$ ($1.5$)\\
& Length-2 & $3.44$ ($1.1$) & $3.50$ ($1.0$) & $3.38$ ($1.3$) & $3.68$ ($1.1$) & $3.33$ ($1.2$) & $4.00$ ($1.1$) & $3.00$ ($1.4$) & $2.50$ ($1.2$) & $3.46$ ($1.5$) & $1.88$ ($1.4$) & $2.17$ ($1.6$) & $1.62$ ($1.1$)\\
& Length-4 & $3.36$ ($1.6$) & $3.42$ ($1.7$) & $3.31$ ($1.5$) & $3.88$ ($1.1$) & $3.67$ ($1.2$) & $4.08$ ($1.0$) & $4.20$ ($1.2$) & $3.58$ ($1.4$) & $4.77$ ($0.4$) & $2.52$ ($1.6$) & $2.42$ ($1.9$) & $2.62$ ($1.3$)\\
\hline
\end{tabular}
}
\caption{Mean (standard deviation) of scores for the Likert-scale items for all participants (\textit{All}) as well as separately for the two conditions (\textit{Learn},\textit{ Credits}).}
\label{tab:diff}
\end{table}


















\section{Discussion}

Existing methods for MOOC recommendation often have a tradeoff between interpretability and accuracy. The original PGPR stands out for being both accurate, comparable to the best neural network-based recommendation approaches, and interpretable, given its capability to generate paths. Nonetheless, its reliance on manually defined path patterns, which require expert knowledge and are not feasible for longer paths, limits its generalizability and real-world application.

Our model maintains the strengths of PGPR and is generalizable to new datasets and Knowledge Graphs. However, prior to deploying UPGPR in real-world settings, several considerations must be addressed.
We have uncovered a nuanced trade-off between accuracy and interoperability. Longer paths boost the model's performance, but our interactions with participants highlighted a preference for shorter, more interpretable paths, corroborating findings from other studies~\cite{DBLP:conf/chi/Kizilcec16} that excessive information can hinder the learning experience. Addressing this trade-off is crucial for real-world deployment and could be achieved through reward engineering (smaller rewards- for longer paths) or by displaying only the shortest path to learners.

Interview insights indicated a favoring of Path-based over Popularity-based explanations, and similar trends were noted with Collaborative Filtering. Moreover, explanations employing ‘Category’ and ‘Teacher’ were favored above ‘Learner’. Given the current KG predominantly features ‘learner enrolled to course’ relations, the prevalent use of these least trusted and least goal-aligned relations by the agent is concerning and could have a negative effect on learners' trust. Mitigating this requires strategic adjustment either at the agent level by promoting the use of diverse relations during training or by enriching the KG with a broader spectrum of categories and concepts. Another potential solution is indicating multiple paths from various learners recommending the same course, diversifying the focus from a single learner, and potentially enhancing trust in the reasoning provided.

In large-scale MOOC platforms, discerning the motivations underlying a student's course selection can be intricate due to the multifaceted learning trajectories they pursue. A single student might opt for one course driven by a genuine interest in the subject matter while concurrently enrolling in another to fulfill their doctoral program requirements. A primary strength of our approach lies in its adaptability to a diverse spectrum of student motivations, objectives, and interests. As evidenced by our findings, offering students a plethora of explanatory paths empowers them to select the recommendation that best aligns with their immediate needs and inclinations.


\section{Conclusion}

In conclusion, our work demonstrates the efficacy of path-based recommendation systems for MOOCs in terms of accuracy and interpretability, a critical aspect in impactful domains like education. Our model is competitive with state-of-the-art, non-explainable models, and its ability to generate understandable path-based explanations over a KG makes it an ideal tool to integrate into transparent and explainable recommendation systems. User preferences, as revealed through our study, lean distinctly towards shorter, more interpretable path-based explanations, emphasizing the importance of avoiding excessive complexity and information overload to maintain user trust and receptivity. While UPGPR represents a step in the right direction for accurate, explainable, and generalizable course recommendation systems, refining the model to address the nuanced preferences and needs of users is crucial for its successful implementation and acceptance in real-world educational settings and beyond.

\bibliographystyle{ACM-Reference-Format}
\bibliography{references}




\end{document}